\documentclass[aps,showkeys,preprintnumbers,amsmath,epsf,amssymb,epsfig,nofootinbib]{revtex4}
\input epsf.sty
\usepackage{bm}
\usepackage{graphicx}
\draft

\usepackage{graphicx}

\newcommand {\eps}{\epsilon}
\newcommand {\ga}{\gamma}
\newcommand {\de}{\delta}
\newcommand {\Ga}{\Gamma}
\newcommand {\La}{\Lambda}
\newcommand {\la}{\lambda}
\newcommand {\al}{\alpha}
\newcommand {\be}{\beta}
\newcommand {\pa}{\partial}
\newcommand {\na}{\nabla}
\newcommand {\fr}{\frac}
\newcommand {\vphi}{\varphi}
\newcommand {\cB}{{\cal B}}
\newcommand {\ca}{{\cal A}}
\newcommand {\ch}{{\cal H}}

\newcommand {\beg}{\begin{equation}}
\newcommand {\en}{\end{equation}}
\newcommand {\bega}{\begin{eqnarray}}
\newcommand {\ena}{\end{eqnarray}}
\begin{document}
\title{Primordial large-scale electromagnetic fields from
Gravitoelectromagnetic Inflation}
\author{ $^{1,2}$ Federico Agust\'{\i}n Membiela \footnote{
E-mail address:membiela@argentina.com}, $^{1,2}$ Mauricio Bellini
\footnote{E-mail address: mbellini@mdp.edu.ar}}
\address{$^{1}$ Departamento de F\'{\i}sica, Facultad de Ciencias Exactas y
Naturales, Universidad Nacional de Mar del Plata, Funes 3350,
(7600) Mar del Plata,
Argentina.\\
$^{2}$ Consejo Nacional de Investigaciones Cient\'{\i}ficas y
T\'ecnicas (CONICET). }

\begin{abstract}
We investigate the origin and evolution of primordial electric and
magnetic fields in the early universe, when the expansion is
governed by a cosmological constant $\Lambda_0$. Using the
gravitoelectromagnetic inflationary formalism with $A_0=0$, we
obtain the power of spectrums for large-scale magnetic fields and
the inflaton field fluctuations during inflation. A very important
fact is that our formalism is {\em naturally non-conformally
invariant}.
\end{abstract}

\maketitle

\section{Introduction}

The origin of the primordial magnetic fields has been subject of a
great amount of research\cite{mag}. The existence, strength and
structure of these fields in the intergalactic plane, within the
Local Superclusted, has been scrutinized recently\cite{ocho}. Many
spiral galaxies are endowed with coherent magnetic fields of $\mu
G$ (micro Gauss) strength~\cite{1,2,3,4,5,6}, having approximately
the same energy density as the Cosmic Microwave Background
Radiation (CMBR). In particular, the field strength of our galaxy
is  $B \simeq 3 \times 10^{-6} \  G$, similar to that detected in
high redshift galaxies~\cite{uno} and damped Lyman alpha
clouds~\cite{dos}. Limits imposed by the high isotropy of CMB
photons, obtained from the COBE data\cite{clarkson} restrict the
present day strength of magnetic fields on cosmological scales to
$10^{-9}\,G$. It is very mysterious that magnetic fields in
clusters of galaxies [i.e., on scales  $ \sim \, {\rm Mpc}$], to
be coherent\cite{camp1}. There are compelling indications of
existence of large-scale microgauss magnetic fields in galaxy
clusters. This would indicate that the entire universe is
magnetized. There are two possible classes of mechanisms to
produce cosmic fields depending on when they are generated:
astrophysical mechanisms acting during large-scale structure
formation, and mechanisms acting in the primordial universe. The
origin of these magnetic fields is not well understood yet. The
seeds of these fields could be in the early inflationary expansion
of the universe, when these fields were originated. The existence
of primordial magnetic fields would affect both, the temperature
and polarization anisotropies of the cosmic microwave background.
It also provides a plausible explanation for the possible
disparity between observations and theoretical fits to the CMB
power-spectrum. The ACBAR\cite{kuo} and CBI\cite{read} experiments
indicate continued power up to $l \sim 4000$, but WMAP data
predicts a rapidly declining power spectrum in the large multipole
range\cite{yama}. This discrepancy is difficult to account from a
returning of cosmological parameters. Among other possible
explanations, an cosmological magnetic field generated during
inflation provides a plausible mechanism to produce excess power
at high multipoles. Therefore, the study of its origin and
evolution in this epoch should be very important to make
predictions in cosmology\cite{giovannini}. During inflation the
extension of the causally connected regions grows as the scale
factor and hence faster than in the decelerated phase. This solves
the horizon problem. Furthermore, during inflation the
contribution of the spatial curvature becomes very small. The way
inflation solves the curvature problem is by producing a very tiny
spatial curvature at the onset of the radiation epoch taking place
right after inflation. The spatial curvature can well grow during
the decelerated phase of expansion but it will be always
subleading provided inflation lasted for sufficiently long time.
It is natural to look for the possibility of generating such a
large-scale magnetic field during inflation. However, the FRW
universe is conformal flat and the Maxwell theory is conformal
invariant, so that magnetic field generated at inflation would
come vanishingly small. Therefore, the conformal invariance must
be broken to generate non-trivial magnetic fields. Various
conformal symmetry breaking mechanisms have been proposed so
far\cite{varios}. Magnetogenesis has been studied also during the
electroweak phase transition\cite{otras}. Due to this fact we are
interested to study a theory that for low energies, we shall
assume reduces to the Maxwell one in the limit of small fields.

Gravitoelectromagnetic Inflation (GI) was developed very recently
with the aim to describe, in an unified manner, the inflaton,
gravitatory and electromagnetic fields during
inflation\cite{g1,g2}. In this formalism all the 4D sources have a
geometrical origin. This formalism can explain the origin of seed
magnetic fields on cosmological scales observed today. This
proposal was constructed from a 5D vacuum state on a
$R^{A}\,_{BCD}=0$ globally flat metric. As in all Space Time
Matter (STM) models\cite{.}, the 4D sources are geometrically
induced when we take a foliation on the fifth coordinate which is
spacelike and noncompact. However, in the previous works was used
the Feynman gauge in order to simplify the structure of the field
equations.

In this letter we shall use this formalism using $A_0=0$. As we
shall see, the field equations become coupled, which has
interesting physical consequences. We shall study the origin and
evolution of the seed large-scale electric and magnetic fields in
a $\Lambda_0$ dominated early universe, from a 5D vacuum state,
where
the expansion of the universe is driven by the inflaton field.  \\

\section{Vector fields in 5D vacuum}

We begin considering a 5D manifold $\cal{M}$ described by a
symmetric metric $g_{AB}=g_{BA}$\footnote{In our conventions
capital Latin indices run from $0$ to $4$, greek indices run from
$0$ to $3$ and latin indices run from $1$ to $3$.}. This manifold
$\cal M$ is mapped by coordinates $\{x^A\}$.
\beg
    dS^2=g_{AB}dx^A dx^B,\label{meta}
\en
which, we shall consider as Riemann-flat $R^A_{BCD}=0$. To
introduce
 the fields we can define an action in $\cal M$.
 \beg
    \mathcal{S}=\int d^5x\sqrt{-g}\left[\fr{^{(5)}\,R}{16\pi
    G}-\fr{1}{4}F_{BC}F^{BC}\right],
 \en
$^{(5)}\,R$ is the 5D scalar curvature. We shall consider these
fields as minimally coupled to gravity. In this space the fields are
free of interactions. The Faraday tensor is antisymmetric
$F_{BC}=\na_B \ca_C-\na_C \ca_B$.

\subsection{The 5D Riemann-flat metric with decaying parameter}

In particular, in this letter we are interested to deal with the
following Riemann-flat metric\cite{coscos}
\beg
    dS^2=\psi^2\fr{\La(t)}{3}dt^2-\psi^2
    e^{2\int^t_0 d\tau\sqrt{\La(\tau)/3}}dr^2-d\psi^2, \label{met1}
\en where $dr^2=dx^i\de_{ij}dx^j$ is the euclidean line element in
cartesian coordinates and $\psi$ is the space-like extra
dimension. Adopting natural units ($\hbar=c=1$) the cosmological
parameter $\La(t)$ (with $\dot\Lambda <0$), has units of
$(length)^{-2}$. The metric (\ref{met1}) is very interesting to
study the evolution of the gravitoelectromagnetic (vectorial)
field, because is Riemann-flat, but has some connections
$\Gamma^{C}_{\,DE}\neq 0$. This fact is very important when we
consider the covariant derivative of $A^F$.

The equations of motion for the components of the vectorial field
${\cal A}$, are

\begin{eqnarray}
\frac{\partial^2 A_4}{\partial t^2} &+& \left[
3\sqrt{\frac{\Lambda}{3}}-\frac{\dot\Lambda}{2\Lambda}\right]
\frac{\partial A_4}{\partial t} - \frac{\Lambda}{3} e^{-2\int
\sqrt{\frac{\Lambda}{3}} dt}\, \nabla^2 A_4 + \frac{\Lambda}{3}
e^{-2\int \sqrt{\frac{\Lambda}{3}}
dt}\,\frac{\partial}{\partial\psi}\left(\vec\nabla .
\vec{A}\right)=0,  \label{aa1} \\
\frac{\partial}{\partial t} \left(\vec{\nabla}.\vec{A}\right)
&+&\psi^2\,e^{2\int \sqrt{\frac{\Lambda}{3}} dt}
\,\frac{\partial}{\partial t}\left(\frac{\partial
A_4}{\partial\psi}\right)
+2\psi\,e^{2\int \sqrt{\frac{\Lambda}{3}} dt}\, \frac{\partial A_4}{\partial t} =0 , \label{aa2} \\
\frac{\partial^2 A_i}{\partial t^2} &+& \left[
\sqrt{\frac{\Lambda}{3}}-\frac{\dot\Lambda}{2\Lambda}\right]
\frac{\partial A_i}{\partial t} - \frac{\Lambda}{3} \,e^{-2\int
\sqrt{\frac{\Lambda}{3}} dt}\, \nabla^2 A_i - \fr{\La}{3}\psi^2
\left[\frac{\partial ^2 A_i}{\partial \psi^2} + \frac{2}{\psi}
\frac{\partial
A_i}{\partial\psi}\right] + 2\fr{\La}{3}\psi \frac{\partial A_4}{\partial x^i} \nonumber \\
&+& \,\fr{\La}{3}e^{-2\int \sqrt{\frac{\Lambda}{3}} dt}\,
\frac{\partial }{\partial x^i} \left(\vec{\nabla} . \vec{A}\right)
+\fr{\La}{3} \psi^2 \, \frac{\partial}{\partial x^i}
\left(\frac{\partial A_4}{\partial\psi}\right) =0. \label{aa3}
\end{eqnarray}
These are our equations of motion on the metric (\ref{met1}), once
we consider the gauge $A^0=0$. To solve these equations we can
begin considering $\vec{\na}\cdot\vec{A}\equiv f(t,\vec{x},\psi),$
and $A_4\equiv \varphi(t,\vec{x},\psi)$. Next, we make Fourier
transforma in eqs. (\ref{aa1}) and (\ref{aa2}), and use separation
of variables for both, $f_k(t,\psi)\sim F_1(t)F_2(\psi)$ and
$\varphi_k(t,\psi)\sim \al(t)\be(\psi)$ -where we drop de subindex
$k$ for the transformed variables $F_1,F_2,\al,\be$. Working out
(\ref{aa1}) and (\ref{aa2}), we arrive to

 \bega
    b^2\dot{\al}&=&\fr{1}{\la_1}\dot{F}_1,\label{bb1} \\
    \ddot{\al}-\Ga\dot{\al}+k^2\fr{\La}{3b^2}\al&=&\la_2\fr{\La}{3b^2}F_1\label{bb3},\\
    \psi^2\be'+2\psi\be&=&\la_1 F_2,\label{bb2}\\
    -\be&=&\fr{1}{\la_2}F'_2\label{bb4},
  \ena
where primes and dots denote respectively the derivatives with
respect to $\psi$ and $t$. Furthermore, we define $b(t)\equiv
e^{\int d\tau\sqrt{{\La\over 3}}}$ and $\Ga(t)\equiv
\fr{\dot{\La}}{2\La}-\sqrt{3\La}$. The constants $\la_1 , \la_2$
come from the separations of variables procedure. To obtain $\al$
and $F_1$, we work with the equations (\ref{bb1}) and (\ref{bb3}).
We introduce $\sigma\equiv \dot\al$ and we have \beg
    \left[\fr{3b^2}{\La}\right]\ddot{\sigma}+\left[\fr{d}{dt}\left(\fr{3b^2}{\La}\right)-
    \fr{3b^2}{\La}\Ga\right]\dot{\sigma}+\left[k^2-\fr{d}{dt}\left(\fr{3b^2}{\La}\Ga-
    \la_1\la_2 b^2\right)\right]\sigma=0,\label{c1}
\en where the solutions are given by the primitives
\bega
    \al(t)&=&\int \sigma(t)dt, \\
    F_1(t)&=&\la_1\int \sigma(t)b(t)^2 dt.
\ena
To solve the equation for $F_2$ we replace (\ref{bb4}) and
its derivative in (\ref{bb2})
\beg
    \psi^2F''_2+2\psi F'_2+\la_1\la_2=0.
\en
The solution is
\begin{eqnarray}
F_2(\psi) &= &
\psi^{-\fr{1}{2}}\left[c_1\psi^w+c_2\psi^{-w}\right], \\
\be(\psi) &= &
\psi^{-\fr{3}{2}}\left[c_1\left(\fr{\fr{1}{2}-w}{\la_2}\right)
\psi^w+c_2\left(\fr{w+\fr{1}{2}}{\la_2}\right)\psi^{-w}\right],
\end{eqnarray}
with $w\equiv\sqrt{\fr{1}{4}-\la_1\la_2}$. In order to illustrate
the formalism we can study an example, which is interesting for
the cosmological expansion of the early universe.\\

\section{An example with $\Lambda = \Lambda_0$: de Sitter expansion}

We consider the case where $\Lambda = \Lambda_0$. When we make the
foliation $\psi=\psi_0=\sqrt{{3\over\Lambda_0}}=H_0^{-1}$ on the
(\ref{met1}), this case give us a de Sitter inflationary expansion
of the universe with tetra-velocities: $u^{\alpha} = (1,0,0,0)$
for a comoving frame. Furthermore, the effective 4D line element
is
\begin{equation}\label{sitter}
ds^2 = dt^2 - H_0^{-2} \,e^{2 H_0 t} d{r }^2.
\end{equation}
In this case equation (\ref{c1}) yields

 \beg
    \ddot{\sigma}+5H_0\dot{\sigma}+H_0^2\left(k^2e^{-2H_0t}+6-\la_1\la_2\right)\sigma=0,
 \en
\\
which has the solution $\sigma(t)=e^{-\fr{5}{2}H_0t}\left[N_1
J_{-\mu}\left(ke^{-H_0t}\right)+N_2
Y_{-\mu}\left(ke^{-H_0t}\right)\right]$, with $\mu\equiv
\sqrt{\fr{1}{4}+\la_1\la_2}$. To get $\al(t)$ we have to integrate
the Bessel functions with the exponential, this can be achieved by
changing variables to $\eta=-\fr{5}{2}H_0\int dt
e^{-\fr{5}{2}H_0t}=e^{-\fr{5}{2}H_0t}$. The primitive of a first
kind Bessel function gives a Regularized Hypergeometric Function
(RHF), and for the second kind Bessel function we obtain the
solution as a combination of two RHFs
 \bega\label{cc1}
    \al(\eta)=&-&\fr{1}{H_0}M_1 2^{\mu-1}k^{-\mu}\eta^{1-\fr{2}{5}\mu}
    \Ga_{\left(\fr{5-2\mu}{4}\right)}\ {}_1\tilde{F}_2
    \left[\left\{\fr{5-2\mu}{4}\right\};\left\{1-\mu,\fr{9-2\mu}{4}\right\};
    -\fr{k^2}{4}\eta^{4/5}\right]\\
    \nonumber &-&\fr{1}{H_0}M_2 2^{-(\mu+1)} k^{\mu}\eta^{1+\fr{2}{5}\mu}\Ga_{\left(\fr{5+2\mu}{4}\right)}
    {}_1\tilde{F}_2\left[\left\{\fr{5+2\mu}{4}\right\};\left\{1+\mu,\fr{9+2\mu}{4}\right\};
    -\fr{k^2}{4}\eta^{4/5}\right],
 \ena
where $M_1\equiv N_1-\fr{N_2}{tan(\mu\pi)}$ and $M_2\equiv N_2$.
To calculate $F_1$ we can define $\eta'=-\fr{1}{2}H_0\int dt
e^{-\fr{1}{2}H_0t}=e^{-\fr{1}{2}H_0t}$
\bega\label{cc2}
\fr{1}{\la_1}F_1(\eta')=&-&\fr{1}{H_0}M_1
2^{\mu-1}k^{-\mu}\eta'^{1-2\mu} \Ga_{\left(\fr{1-2\mu}{4}\right)}\
{}_1\tilde{F}_2
\left[\left\{\fr{1-2\mu}{4}\right\};\left\{1-\mu,\fr{5-2\mu}{4}\right\};
-\fr{k^2}{4}\eta'^{4}\right]\\
\nonumber &-&\fr{1}{H_0}M_2
2^{-(\mu+1)}k^{\mu}\eta'^{1+2\mu}\Ga_{\left(\fr{1+2\mu}{4}\right)}\
{}_1\tilde{F}_2\left[\left\{\fr{1+2\mu}{4}\right\};\left\{1+\mu,\fr{5+2\mu}{4}\right\};
-\fr{k^2}{4}\eta'^{4}\right]. \ena
The total solution
$\varphi_k(t,\psi)=[\phi_k^{(hom)}(t)+\al_k(t)]\be(\psi)$, where
we have included the homogeneous solution \beg \label{homo}
\phi_k^{(hom)}(t)=A_1 e^{-\fr{3}{2}H_0t} \ch_{3/2}^{(1)} \left(k
e^{-H_0t}\right)+A_2 e^{-\fr{3}{2}H_0t} \ch_{3/2}^{(2)}\left(k
e^{-H_0t}\right), \en that has the typical scale invariant
spectrum of a de Sitter model.

Once obtained the solutions for $\varphi_k(t,\psi)$ and
$f_k(t,\psi)$, we can try to solve the equations for the potential
3-vector $A_j(x^A)$. We take the Fourier transform in the
$\vec{x}$-space, as before. From eq. (\ref{aa3}), we obtain the
following equation for the modes $\xi_k^{(j)}(t,\psi)$: \beg
\ddot{\xi}_k^{(j)}+H_0\dot{\xi}_k^{(j)}+ H_0^2
k^2e^{-2H_0t}\xi_k^{(j)}-
\psi^2H_0^2\left[\fr{\pa^2}{\pa\psi^2}+\fr{2}{\psi}\fr{\pa}{\pa\psi}\right]\xi_k^{(j)}=
-ik_jH_0^2\underbrace{\left[2\psi\phi_k(t)\be(\psi)+e^{-2H_0t}f_k(t,\psi)+
\psi^2\phi_k(t)\be'(\psi)\right]}_{{\cal K}(t,\psi)}, \label{mell}
\en where ${\cal K}(t,\psi)$ is the source term and
\begin{equation}\label{extra}
\phi_k(t)=\phi_k^{(hom)}(t)+\alpha_k(t).
\end{equation}
In this letter we consider that the particle excitations for an
observer in (\ref{met1}) appear are the Mellin transform in the
extra coordinate $\psi$. We will see that the extra terms will
become massive terms for each $m$-mode. Thus, this extra
coordinate formalism, besides it produces couplings between the
effective vector and scalar components of the field (in a curved
background), provide us of a contribution to the mass of vector
excitations. The Mellin transform on the foliated spacetime
(\ref{sitter}), is
 \beg
    \xi_{k,m}^{(j)}(t)=\int_0^{1}\psi'^{m-1}\xi_k^{(j)}(t,\psi')d\psi',\
    \ \  \psi'=\fr{\psi}{\psi_0},
 \en
\\
and the equation (\ref{mell}) becomes
 \beg\label{cc3}
    \ddot{\xi}_{k,m}^{(j)}+H_0\dot{\xi}_{k,m}^{(j)}+H_0^2\left[ k^2e^{-2H_0t}
    -m(m-1)\right]\xi_{k,m}^{(j)}=-ik_jH_0^2{\cal K}_m(t),
 \en
where $m$ is a free parameter on the metric (\ref{sitter}) [but
not on the 5D Riemann-flat metric (\ref{met1})], to be
experimentally determined by the spectrum of large scale magnetic
fields. The total solution to this ordinary differential equation
is the homogenous part plus the inhomogeneous one
\begin{equation}\label{cc4}
    \xi_{k,m}^{(j)}(t)=D_1
    e^{-\fr{1}{2}H_0t}\ch_{\fr{1}{2}-m}^{(1)}\left[x(t)\right]+
    D_2 e^{-\fr{1}{2}H_0t}\ch_{\fr{1}{2}-m}^{(2)}\left[x(t)\right]
    + \left.\xi_{k,m}^{(j)}\right|_{inh},
\end{equation}
where
\begin{equation}
\left.\xi_{k,m}^{(j)}\right|_{inh}= ik_jH_0\fr{\pi}{2}
e^{-\fr{1}{2}H_0t}\int d\tau
    {\cal
    K}_m(\tau)e^{\fr{1}{2}H_0\tau}\left[Y_{\fr{1}{2}-m}\left[x(\tau)\right]J_{\fr{1}{2}-m}\left[x(t)\right]+
    J_{\fr{1}{2}-m}\left[x(\tau)\right]Y_{\fr{1}{2}-m}\left[x(t)\right]\right].\label{inh}
\end{equation}
The source term, after the Mellin transform, yields
 \beg
    {\cal K}_m(\tau)=\fr{c_1}{\left(m-\fr{1}{2}+w\right)}\left[e^{-2H_0\tau}F_1(\tau)+
    \fr{\left(\fr{1}{2}+w\right)^2}{\la_2}\al(\tau)\right]+
    \fr{c_2}{\left(m-\fr{1}{2}-w\right)}\left[e^{-2H_0\tau}F_1(\tau)+
    \fr{\left(\fr{1}{2}-w\right)^2}{\la_2}\al(\tau)\right].
 \en
Notice that $F_1(\tau)$ and $\al(\tau)$ are given by
hypergeometric functions and we have to integrate them with the
Bessel functions. This can be done analytically by evaluating the
integral for each term of the power series of the RHF. To make it,
we write explicitly the RHFs in the form

 \beg \label{hyper}
    {}_1\tilde{F}_2\left[\{a\};\{b_1,b_2\};z\right]=\sum_{p=0}^\infty\ga_p
    \fr{z^p}{z!},
 \en
with
$\ga_p\equiv\fr{\Ga_{(a+p)}}{\Ga_{(b_1+p)}\Ga_{(b_2+p)}\Ga_{(p)}}$.
The integrals we have to evaluate are of the form
 \beg
    \sum_{p=0}^\infty\fr{\ga_p\left(-\fr{k^2}{4}\right)^p}{p!}\int d\tau
    \left[e
    ^{-2pH_0\tau}K_{\fr{1}{2}-m}\left(ke^{-H_0\tau}\right)\right]e^{-\left(\pm\mu+\fr{3}{2}\right)H_0\tau},\label{int}
 \en
where $K_{\fr{1}{2}-m}[x(t)]$ can be either the first kind or
second kind Bessel function. To solve the integral again we repeat
the procedure of changing variables
$\eta''_{\pm}=-\left(\pm\mu+\fr{3}{2}\right)\int d\tau
e^{-\left(\pm\mu+\fr{3}{2}\right)H_0\tau}=e^{-\left(\pm\mu+\fr{3}{2}\right)H_0\tau}$,
so the integrals in (\ref{int}) reduce to $\int d\eta''_{\pm}
{\eta''_{\pm}}^{\fr{2p}{\pm\mu+\fr{3}{2}}}K_{\fr{1}{2}-m}\left[k
{\eta''_{\pm}}^{\fr{1}{\pm\mu+\fr{3}{2}}}\right]$. The result for
the inhomogeneous solution (\ref{inh}), is
 \bega
    \left.\xi_{k,m}^{(j)}\right|_{inh}& =& -\sum_{s=1,2}^{} \sum_{n=1,2}^{} \fr{i\pi M_s c_n }{2^{2+\mu_s}H_0
    \left(m+w_n-\fr{1}{2}\right)}k_j e^{-2H_0t}
    \left(k e^{-H_0t}\right)^{\mu_s} \nonumber \\
    &\times &
    \left\{\la_1\left[J_{\fr{1}{2}-m}[x(t)]\mathbb{I}_2^{(1)}(t)+
    Y_{\fr{1}{2}-m}[x(t)]\mathbb{I}_1^{(1)}(t)\right]+\fr{\left(w_n+\fr{1}{2}\right)^2}{\la_2}
    \left[J_{\fr{1}{2}-m}[x(t)]\mathbb{I}_2^{(2)}(t)+
    Y_{\fr{1}{2}-m}[x(t)]\mathbb{I}_1^{(2)}(t)\right]\right\},
 \ena
with
 \bega
         \mathbb{I}_1^{(r)}(t)&=&\sum_{p,q=0}^\infty\ga_p^{(r)}\ga_q^{(2)}
         \fr{\left(-\fr{k e^{-H_0t}}{2}\right)^{2(p+q)}}{p!q!}
         2^{m-2p-\fr{3}{2}}\left(k e^{-H_0t}\right)^{\fr{1}{2}-m},\ \ \ \ r=1,2 \label{31}\\
         \mathbb{I}_2^{(r)}(t)&=&\sum_{p,q=0}^\infty\ga_p^{(r)}
         \fr{\left(-\fr{ke^{-H_0t}}{2}\right)^{2(p+q)}}{p!q!}
         2^{m-2p-\fr{3}{2}}\left(k e^{-H_0t}\right)^{\fr{1}{2}-m}
         \left[-2 {\rm sec}(m\pi)\left(k
         e^{-H_0t}\right)^{2m}\ga_q^{(1)}\right.
\nonumber \\
&+&\left.{\rm tan}(m\pi)4^m k
e^{-H_0t}\ga_q^{(2)}\right], \label{32} \ \ \ \ r=1,2,
 \ena
and the coefficients $\ga_p$ and $\ga_q$, are
 \bega
    \ga_p^{(1)}&=&\fr{\Ga_{\left(\fr{1+2\mu_n}{4}\right)}}{\Ga_{(p)}\Ga_{(1+\mu_n+p)}\left(\fr{1+2\mu_n}{4}+p\right)},\\
    \ga_p^{(2)}&=&\fr{\Ga_{\left(\fr{5+2\mu_n}{4}\right)}}{\Ga_{(p)}\Ga_{(1+\mu_n+p)}\left(\fr{5+2\mu_n}{4}+p\right)},\\
    \ga_q^{(1)}&=&\fr{\Ga_{\left(\fr{1+m+\mu_n}{2}+p\right)}}{\Ga_{(q)}\Ga_{(\fr{1}{2}+m+q)}\left(\fr{1+m+\mu_n}{2}+p+q\right)},\\
    \ga_q^{(2)}&=&\fr{\Ga_{\left(1-\fr{m-\mu_n}{2}+p\right)}}{\Ga_{(q)}\Ga_{(\fr{3}{2}-m+q)}\left(1-\fr{m-\mu_n}{2}+p+q\right)},
 \ena
 where $\mu_1=-\mu_2=\mu$ and $w_1=-w_2=w$.

To study the modes of cosmological interest, we have to work with
those modes that during inflation stay outside the horizon. This
is the limit $ke^{-H_0t}\ll 1$. Then we truncate the power series
to
the first term $p=q=0$.\\
We also introduce a constraint in the possible values of $m$. If
we look at equation (\ref{cc3}), we see that, in order to identify
this term with ordinary matter, the quantity $m(m-1)>0$ in eqs.
(\ref{31}) and (\ref{32}). This means that $m<0$ or $m>1$. Then,
to keep positive parameters for the Hankel functions in
(\ref{cc4}), we shall work with the negative values of $m$. Notice
that $\mu$ and $w$ are not independent parameters. They are
related by the equation, $\mu^2+w^2=\fr{1}{2}$. If we require that
$\mu$ and $w$ be real and positive, then we obtain that they are
restricted to the interval $\left[0,\fr{\sqrt{2}}{2}\right]$.
This restricts the parameter space of $\la_1,\la_2$.\\

Considering that the magnetic fields produced during this epoch are
scale invariant in the cosmological level, we can define a value for
the parameter $m$
 \beg\label{d1}
    \cB_{com}^2\equiv \langle B_{com}^2\rangle=\fr{1}{2\pi^2}\int_0^{\theta k_H(t)}\fr{dk}{k}k^{5}
    \xi_{k,m}^{(j)}\xi_{k,m}^{(j)\star}
 \en
The contribution to the magnetic fields is exclusive from the
homogeneous solution in (\ref{cc4}), this should be clear from
(\ref{aa3}) because the source terms are all gradients. The k-power
that come from the homogeneous solution of the modes is
$m-\fr{1}{2}$. Then, to obtain a nearly scale invariant spectrum of
the magnetic field we need that $5-1+2m\simeq0$, yielding
$m\simeq-2$. In the cosmological limit the inhomogeneous equation
can be reduced to
 \bega\label{dd1}
    \left.\xi_{k,-2+\eps}^{(j)}\right|_{inh}& =& -\sum_{n=1,2}^{}\fr{i\pi
    M_1
    c_n}{2^{2-\mu}H_0
    \left(w_n-\fr{5}{2}\right)}k_j e^{-2H_0t}
    \left(k e^{-H_0t}\right)^{-\mu}\fr{\Ga_{\left(\fr{1+2\mu}{4}\right)}}{\Ga_{(1+\mu)}}
    \left\{\fr{4\la_1}{\left(1+2\mu\right)}+
    \fr{\left(w_n+\fr{1}{2}\right)^2}{\la_2}\left(\fr{1+2\mu}{5+2\mu}\right)
    \Ga_{\left(\fr{1+2\mu}{4}\right)}\right\} \nonumber \\
    &\times &
    \left[\left(\fr{\eps}{225}\left(k e^{-H_0t}\right)^{5}-\fr{2^{\fr{3}{2}}}{45\pi}\right)
    \fr{\Ga_{(2+\mu/2)}}{2+\mu/2}-\fr{8\sqrt{\pi}\Ga_{\left(\fr{\mu-1}{2}\right)}}{3(\mu-1)}\right],
 \ena
where $\eps$ is a parameter that takes into account small
deviations from scale invariance. In the last expression should be
noted that we have drop the terms involving positive $\mu$-powers
of the physical wavenumber $ke^{-H_0t}$ (this because of the same
idea we stayed with $p=q=0$). The Bessel functions have been
approximated to their asymptotic expressions in the infrared
limit. Note that the term involving $\eps$ decays very strongly
due to the factor $x(t)^5$.

\subsection{Quantization and normalization of homogeneous solutions of the vector and scalar fields}

We consider the usual commutation relations for the fields and
their conjugate momenta on the effective 4D metric (\ref{sitter}).
For the scalar field we obtain
 \beg
    \left[\vphi_m{(t,\vec{x}),\dot{\vphi}_m(t,\vec{x}')}\right]=i
    H_0^3e^{-3H_0t}\de^{(3)}(\vec{x}-\vec{x}'),
 \en
where $\vphi_m(t,\vec{x})=\vphi(t,\vec{x})\tilde\be_m$, and
$\tilde\be_m$ is the Mellin transform of $\be(\psi)$. From these
relations we derive the normalization condition for the modes
 \beg
    \phi_k\dot\phi_k^{\star}-\dot\phi_k\phi_k^\star=\fr{i}{m^2|\tilde\be_m|^2}H_0^3e^{-3H_0t},
 \en
\\
where the Bunch-Davies vacuum is given by $A_1=0$ and then, we
obtain $A_2=i\fr{H_0\sqrt\pi}{2|m\tilde\be_m|}$. We repeat the
same for the vector solution, the commutation relation is in this
case
 \beg
    \left[{\cal A}_{m,j}(t,\vec{x}),\dot{{\cal
    A}}_{m',k}(t,\vec{x}')\right]=i \de_{jk}\,\de_{mm'}\,
    H_0e^{-H_0t}\de^{(3)}(\vec{x}-\vec{x}'),
 \en
where ${\cal A}_{m,j}(t,\vec{x})$ is the Mellin transform of the
$A_j$ field. We impose the Bunch-Davies vacuum to the modes
choosing $D_1=0$ and $D_2=i\fr{\sqrt{\pi}}{2}$ in (\ref{cc4}).
Then, the Fourier-Mellin modes comply
 \beg
    \xi_{k,m}^{(j)}(t)\dot\xi_{k,m}^{(j)\star}(t)-\xi_{k,m}^{(j)\star}(t)\dot\xi_{k,m}^{(j)}(t)=
    iH_0e^{-H_0t}.
 \en
We have noted that the magnetic fields depend exclusively from the
homogeneous solution, then we can compute their quadratic
amplitude from (\ref{d1})
 \bega
    {\langle B^iB_i\rangle}^{\fr{1}{2}}=H_0 e^{-H_0t}\cB_{com}=
    \fr{\Ga_{(\fr{1}{2}-m)}}{(2\pi)^{3/2}2^m}H_0 e^{-(1+m)H_0t}\fr{(\theta k_H)^{2+m}}{\sqrt{|2+m|}},\\
    \cB_{phys}=H_0^2e^{-2H_0t}\cB_{com}= \fr{\Ga_{(\fr{1}{2}-m)}}{(2\pi)^{3/2}2^m}H_0^2 e^{-(m+2)H_0t}
    \fr{(\theta k_H)^{2+m}}{\sqrt{|2+m|}}.
 \ena
The horizon wavenumber is found to be
$k_H(t)=\sqrt{\fr{1}{4}+m(m-1)}e^{H_0t}.$ This means that the
physical magnetic field is constant for any $m$
 \beg
    \cB_{phys}=H_0^2e^{-2H_0t}\cB_{com}= \fr{\Ga_{(\fr{1}{2}-m)}}{(2\pi)^{3/2}2^m}H_0^2
    \fr{\left[\theta
    \sqrt{\fr{1}{4}+m(m-1)}\right]^{2+m}}{\sqrt{|2+m|}},
 \en
which is divergent for $m=-2$. If we require that these fields are
nearly invariant in cosmological scales, then $m=-2+\eps$, $\eps$
being a small parameter, this yields for the amplitude of the
physical magnetic field
 \beg
    \cB_{phys}=\fr{3}{4}\sqrt{\fr{2}{\pi}}H_0^2\fr{(5\theta/2)^{\eps}}{\sqrt{\eps}}.
 \en
On the other hand the electric fields are affected by the extra
terms. Then, at the end of the inflationary epoch there will be a
contribution from the inhomogeneous solution to the spectrum and
amplitude of the electric field. This field has components
$E^\al=F^{\al\be}v_\be$, where $v_\be$ are the components of the
observer velocity. For a physical observer we have
$\vec{E}_{phys}=a^{-1}(t)\pa_t \vec{A}$. Therefore
 \beg
    {\cal E}^2_{phys}\equiv \langle E_{phys}^2\rangle=\fr{H_0^2e^{-2H_0t}}{2\pi^2}\int
    \fr{dk}{k}k^3 \dot\xi_{k,m}^{(j)} \dot\xi_{k,m}^{(j)\star}.
 \en
From the homogeneous solution product we obtain the amplitude
 \beg
    \langle {E^{[h]}_{phys}}^2\rangle=\fr{H_0^2\,e^{-2H_0t}}{2\pi^2}\int
    \fr{dk}{k}k^3 \left|{\dot\xi_{k,m}^{[h](j)}}\right|^2.
 \en
Hence, the homogeneous amplitude of the electric field, is
 \beg
    {\cal
    E}^{[h]}_{phys}=\fr{\sqrt{2}\,\,\Ga\left[\fr{1}{2}-m\right]}{(2\pi)^{3/2}2^m
    }|m|H_0^2
    \fr{\left[\theta
    \sqrt{\fr{1}{4}+m(m-1)}\right]^{m+3/2}}{\sqrt{|2m+3|}}e^{\fr{1}{2}H_0t}.
 \en
The relation between the energy density of the electric and
magnetic fields, in the physical frame, is
 \beg
    \fr{\rho_{elec}^{[h]}}{\rho_{mag}}=\left(\frac{2m^2}{\theta}\right) \left(\fr{2+m}{ 3+2m}\right)\fr{e^{H_0t}}{\sqrt{\fr{1}{4}+m(m-1)}}.
 \en
For a nearly scale invariant magnetic field, we obtain
 \beg
    \fr{\rho_{elec}^{[h]}}{\rho_{mag}}=\left(\fr{8\eps}{5\theta}\right)\,e^{H_0t}.
 \en
This means that the energy of the electric field is dominant
during exponential inflation on cosmological scales. For a nearly
scale invariant electric field we obtain $m=-3/2+\eps$, and then
 \beg
     \cB_{phys}=\sqrt{\fr{2\theta}{\pi^3}}H_0^2.
 \en
Thus, the magnetic field would be constant but bigger on smaller
(astrophysical) scales, while the electric field would still
dominate in larger scales
 \beg
    \fr{\rho_{elec}^{[h]}}{\rho_{mag}}=\left(\fr{9}{8\eps\theta}\right)\,e^{H_0t}.
 \en
\subsection{Spectrum of the inhomogeneous solutions of the electric field and the scalar}

The amplitude and spectrum of the electric field have terms that
involve the inhomogeneous solution, with double infinitum power
series, coming from an integral of the hypergeometric functions.
For convenience we only keep the first term $(p=q=0)$, because it
is of cosmological relevance. The contribution of terms with $p$
and $q $ $\neq 0$ can be neglected on cosmological scales. The
three remaining contributions are
 \bega
     \langle {E^{[1]}_{phys}}^2\rangle&=&\fr{H_0^2e^{-2H_0t}}{2\pi^2}\int
    \fr{dk}{k}k^3
    {\dot\xi_{k,m}^{[h](j)}}{\dot\xi_{k,m}^{[inh](j)\star}}, \label{el1}\\
    \langle {E^{[2]}_{phys}}^2\rangle&=&\fr{H_0^2e^{-2H_0t}}{2\pi^2}\int
    \fr{dk}{k}k^3
    {\dot\xi_{k,m}^{[inh](j)}}{\dot\xi_{k,m}^{[h](j)\star}}, \label{el2}\\
    \langle {E^{[3]}_{phys}}^2\rangle&=&\fr{H_0^2e^{-2H_0t}}{2\pi^2}\int
    \fr{dk}{k}k^3
    {\dot\xi_{k,m}^{[inh](j)}}{\dot\xi_{k,m}^{[inh](j)\star}}. \label{el3}
 \ena
In order to simplify the notation, we write (\ref{dd1}) in the
compact form

 \beg
    \xi_{k,-2+\eps}^{[inh](j)}(t)=-\fr{i\,
    e_j}{H_0}e^{-H_0t}\left(k\,e^{-H_0t}\right)^{1-\mu}\sum_{n=1,2}^{}\mathbb{D}_{\la_1\la_2}^{(n)},
 \en
where we used $k_j=k \,e_j$ and the coefficients
$\mathbb{D}_{\la_1\la_2}^{(n)}$ have units of $H_0$. This yields
the respective power spectrums, for (\ref{el1}), (\ref{el2}) and
(\ref{el3})

 \bega
    \langle {E^{[1]}_{phys}}^2\rangle \sim \langle {E^{[2]}_{phys}}^2\rangle \sim \Large{\int} \frac{dk}{k} \,k^{\fr{3}{2}-\mu},\\
    \langle {E^{[3]}_{phys}}^2\rangle \sim \Large{\int} \frac{dk}{k}\, k^{5-2\mu},
 \ena
where we have considered only terms with $p=q=0$ in eqs.
(\ref{31}) and (\ref{32}). It is important to notice that the
spherical symmetry is broken because $e_j$ is an unitary vector.

On the other hand, the power spectrum for the homogeneous part of
the scalar field is scale invariant for the solution
(\ref{extra}), while that for the other terms in (\ref{extra}) we
obtain two different spectrums to zero order in the hypergeometric
function (\ref{hyper})
 \bega
    \langle {\phi^{[1]}}^2 \rangle=\fr{1}{2\pi^2}\int
    \fr{dk}{k}k^3\, \phi^{(hom)}_k(t)\al^{\star}_k(t)\sim e^{-(5/2-\mu)}\int\fr{dk}{k}\, k^{\fr{3}{2}-\mu},\\
    \langle {\phi^{[2]}}^2 \rangle=\fr{1}{2\pi^2}\int
    \fr{dk}{k}k^3\, \left[\phi^{\star}\right]^{(hom)}_k(t)\al_k(t)\sim e^{-(5/2-\mu)}\int\fr{dk}{k}\, k^{\fr{3}{2}-\mu},\\
    \langle {\phi^{[3]}}^2 \rangle=\fr{1}{2\pi^2}\int
    \fr{dk}{k}k^3\, \al_k(t)\al^{\star}_k(t)\sim e^{-2(5/2-\mu)}\int\fr{dk}{k}\, k^{3-2\mu}.
     \ena
Notice that these inhomogeneous terms are exponentially damped.
The parameter $\mu$ can be fixed, with the smaller index that
decays weaker, so as to yield the experimental data for the scalar
spectral index \cite{PDG}, $n_s=0.958$, then we write
$3/2-\mu=\eps'$, with $\eps'=n_s-1\sim -0.042$.

 \section{Final Comments}

In this letter we have studied the primordial spectrum of
electromagnetic fields using GI. Starting from a gauge with
$A_0=0$, we have obtained some interesting properties. In the
example here studied the spectrum of large-scale magnetic fields
is nearly scale-invariant for $m\simeq -2$. The amplitude for the
strength of comoving magnetic fields is dramatically increasing,
but they are frozen in physical coordinates. The important result
here obtained is that the modes of $A_j$ are affected by a source,
which is originated in the modes of the inflaton field, so that
the spectrum of the large-scale electric field during inflation
depends of the modes of the inflaton field. These modes can be
considered as massive photons which are gauge-invariant in a 5D
sense, but once the foliation $d\psi=0$ is done (which implies the
choice of a relativistic system), these photons acquire mass
because they live in an effective 4D curved spacetime. In this
sense the choice of the relativistic system acts as an effective
Higgs's mechanism.

But the more interesting result relies in that the spectrum of the
inflaton field depends of the modes of $A_i$, because they are
coupled to the modes of $A_i$. Of course, this scale invariance is
significatively affected on shorten scales, so that it is nearly
scale invariant on very large scales. This result disagrees with
standard 4D versions of inflation, but it agree very much with
experience, because it is very known that for shorten scales the
mass spectrum of matter has a positive index with a scale
dependent power.

The effects of a conducting plasma in the early inflationary
universe are negligible. During inflation conformal invariance is
broken and the strength of comoving magnetic fields increases
dramatically as $a^2$, until values of the order of ${\cal
B}_{com} \simeq 10^{127}\,{\rm Gauss}$ after $63$ $e-$folds, so
that the flatness problem is resolved in the model. After
inflation, the universe enters in the so-called reheating phase,
during which the energy of the inflaton is converted into ordinary
matter. In this epoch, the conductivity $\sigma_c$ of the universe
is of the order of $\sigma_c \sim T \gg H$ (with a background
temperature $T \ll M_p$). magnetic fields evolves adiabatically
from the end of inflation until today, due to the high electrical
conductivity of the cosmic plasma. In this epoch the universe is
thermalized, so that the comoving magnetic field decreases with
the expansion to take actual values of the order of $10^{-9}\,
{\rm Gauss}$\cite{cobe,g1}. Notice that the results here obtained
depends on the gauge $A_0=0$. It is well known that any viable
mechanism to generate seed magnetic fields during inflation must
repose on the breaking of conformal invariance of standard
electrodynamics. Otherwise, the produced fields are vanishingly
small. Notice that the approach here worded is not conformally
invariant on the effective 4D metric (\ref{sitter}). The origin of
this rupture is in the fact that some connections $\Gamma^C_{DE}$
are non-zero on the 5D Riemann flat metric (\ref{met1}). This is
the reason by which bosons are massive on the effective 4D
spacetime (\ref{sitter}) on which move the observers. Concerning
electric fields, there is a damping of the longitudinal component
of the field strength, corresponding to the gradual neutralization
of charged particles in the primordial plasma\cite{ult}, in the
first stages of reheating. Finally, in our model, inflation occurs
at a very low scale with $H \sim 10^{-9} \, M_p$ and with the
inflaton field taking values much below of the Planckian scale:
$\left< \varphi\right> \simeq 10^{-12}\,M_p$\cite{mb}. In this
sense our model evolves on scales similar to the MSSM inflationary
model\cite{mssm}, where fine tunning and slow rolling problems
joined with reheating were considered and the inflaton field
couplings to Standard Model physics is explained from first
principles. In our case the couplings between the fields $A_C$ is
explained from the induced curvature of the metric (\ref{sitter}).
The problem of back-reaction\cite{ultimo} should be considered in
future works.

\begin{acknowledgments}
The authors acknowledge CONICET and UNMdP (Argentina) for
financial support.
\end{acknowledgments}

\end{document}